\documentclass[runningheads]{llncs}
\usepackage{graphicx}
\usepackage{subfig}
\usepackage{xkeyval,xcolor}% http://ctan.org/pkg/{xkeyval,xcolor}
\makeatletter
\newlength{\sfp@hseplen}\newlength{\sfp@vseplen}
\define@cmdkey{subfigpos}[sfp@]{pos}[ul]{}% \sfp@pos
\define@cmdkey{subfigpos}[sfp@]{font}[\small]{}% \sfp@font
\define@cmdkey{subfigpos}[sfp@]{vsep}[1\baselineskip]{\setlength{\sfp@vseplen}{\sfp@vsep}}% \sfp@vsep
\define@cmdkey{subfigpos}[sfp@]{hsep}[10pt]{\setlength{\sfp@hseplen}{\sfp@hsep}}% \sfp@hsep
\newcommand{\subfigimg}[3][,]{%
  \setkeys{Gin,subfigpos}{pos,font,vsep,hsep,#1}% Set (default) keys
  \setbox1=\hbox{\includegraphics{#3}}% Store image in box
  \ifnum\pdfstrcmp{\sfp@pos}{ul}=0% UPPER LEFT placement of subfig label
    \leavevmode\rlap{\usebox1}% Print image
    \rlap{\hspace*{\sfp@hsep}\raisebox{\dimexpr\ht1-\sfp@vsep}{\sfp@font{#2}}}% Print label
    \phantom{\usebox1}% Insert appropriate spacing
  \else\ifnum\pdfstrcmp{\sfp@pos}{ur}=0% UPPER RIGHT placement of subfig label
    \leavevmode\usebox1% Print image
    \llap{\raisebox{\dimexpr\ht1-\sfp@vsep}{\sfp@font{#2}}\hspace*{\sfp@hsep}}% Print label
  \else\ifnum\pdfstrcmp{\sfp@pos}{lr}=0% LOWER RIGHT placement of subfig label
    \leavevmode\usebox1% Print image
    \llap{\raisebox{\sfp@vsep}{\sfp@font{#2}}\hspace*{\sfp@hsep}}% Print label
  \else% Assume LOWER LEFT placement of subfig label
    \leavevmode\rlap{\usebox1}% Print image
    \rlap{\hspace*{\sfp@hseplen}\raisebox{\sfp@vsep}{\sfp@font{#2}}}% Print label
    \phantom{\usebox1}% Insert appropriate spacing
  \fi\fi\fi
}
\makeatother
\begin{document}
\title{Deep Small Bowel Segmentation with Cylindrical Topological Constraints}
\titlerunning{Small Bowel Segmentation with Topological Constraints}
% If the paper title is too long for the running head, you can set
% an abbreviated paper title here
%
\author{
Seung Yeon Shin\inst{1} \and
Sungwon Lee\inst{1} \and
Daniel C. Elton\inst{1} \and
James L. Gulley\inst{2} \and
Ronald M. Summers\inst{1}
}
%1{Shin, Seung Yeon}
%2{Lee, Sungwon}
%3{Elton, Daniel}
%4{Gulley, James}
%5{Summers, Ronald}
%
\authorrunning{S.Y. Shin et al.}
% First names are abbreviated in the running head.
% If there are more than two authors, 'et al.' is used.
%
\institute{Imaging Biomarkers and Computer-Aided Diagnosis Laboratory, Radiology and Imaging Sciences, Clinical Center, National Institutes of Health, Bethesda, MD, USA\\
\email{\{seungyeon.shin,rms\}@nih.gov} \and
Center for Cancer Research, National Cancer Institute, National Institutes of Health, Bethesda, MD, USA
}

%\institute{Princeton University, Princeton NJ 08544, USA \and
%Springer Heidelberg, Tiergartenstr. 17, 69121 Heidelberg, Germany
%\email{lncs@springer.com}\\
%\url{http://www.springer.com/gp/computer-science/lncs} \and
%ABC Institute, Rupert-Karls-University Heidelberg, Heidelberg, Germany\\
%\email{\{abc,lncs\}@uni-heidelberg.de}}
%
\maketitle              % typeset the header of the contribution
\begin{abstract}
We present a novel method for small bowel segmentation where a cylindrical topological constraint based on persistent homology is applied. To address the touching issue which could break the applied constraint, we propose to augment a network with an additional branch to predict an inner cylinder of the small bowel. Since the inner cylinder is free of the touching issue, a cylindrical shape constraint applied on this augmented branch guides the network to generate a topologically correct segmentation. For strict evaluation, we achieved an abdominal computed tomography dataset with dense segmentation ground-truths. The proposed method showed clear improvements in terms of four different metrics compared to the baseline method, and also showed the statistical significance from a paired t-test.

\end{abstract}
\section{Introduction}

% small bowel, small bowel in CT
The small bowel is the longest (20 $\sim$ 30 ft) section of the digestive tract. While surrounded by other organs including the large bowel, it is pliable and has many folds which allow it to fit into the abdominal cavity~\cite{cc19}. Computed tomography (CT) has become a primary imaging technique for small bowel disease diagnosis among others such as enteroclysis and endoscopy due to its diagnostic efficacy and efficiency~\cite{murphy14}. Apart from its convenience, acquired 3D CT scans are examined by radiologists slice-by-slice in the interpretative procedure, which is very time-consuming. Also, this may be error-prone since the small bowel is close to other abdominal organs such as the large bowel and muscle, both in position and appearance.

% previous work (small bowel segmentation)
Automatic segmentation of the small bowel may improve the procedure and thus could help precise localization of diseases and preoperative planning by better visualization. There have been only a few previous works on automatic small bowel segmentation~\cite{zhang13,oda20}. In \cite{zhang13}, the anatomic relationship between the mesenteric vasculature and the small bowel is used to guide the small bowel segmentation. A limitation of this method is that it requires a CT scan done in the arterial phase. Therefore, it would not work on routine contrast-enhanced CT scans which are done during the portal venous phase. Recently, in \cite{oda20}, the 3D U-Net~\cite{cicek16} is trained for small bowel segmentation with sparsely annotated CT volumes (Seven axial slices for each volume) to avoid making dense annotation.

% previous work (persistent homology)
In the past decade, attempts to utilize topological features in tasks such as disease characterization~\cite{adcock14,belchi18}, organ segmentation~\cite{segonne15,clough19}, and neuron image segmentation~\cite{hu19} have been made. Persistent homology (PH) is a method to compute such topological features underlying in a space~\cite{otter17}. It measures the persistencies of topological features as some threshold changes, and thus represents the robustness of each feature. Based on its differentiable property~\cite{bruel19}, PH is also incorporated into the training process of neural networks~\cite{clough19,hu19}. In \cite{clough19}, the method is applied to segmenting two anatomical structures, which are the myocardium of the left ventricle of the heart from short-axis view 2D cardiac magnetic resonance images and the placenta from 3D ultrasound volumes. Their topologies are clearly defined as ring-shaped and one single connected component with no loops or cavities, respectively.

\begin{figure}[t]
	\centering
	\subfloat[]{\includegraphics[width = 0.4\textwidth]{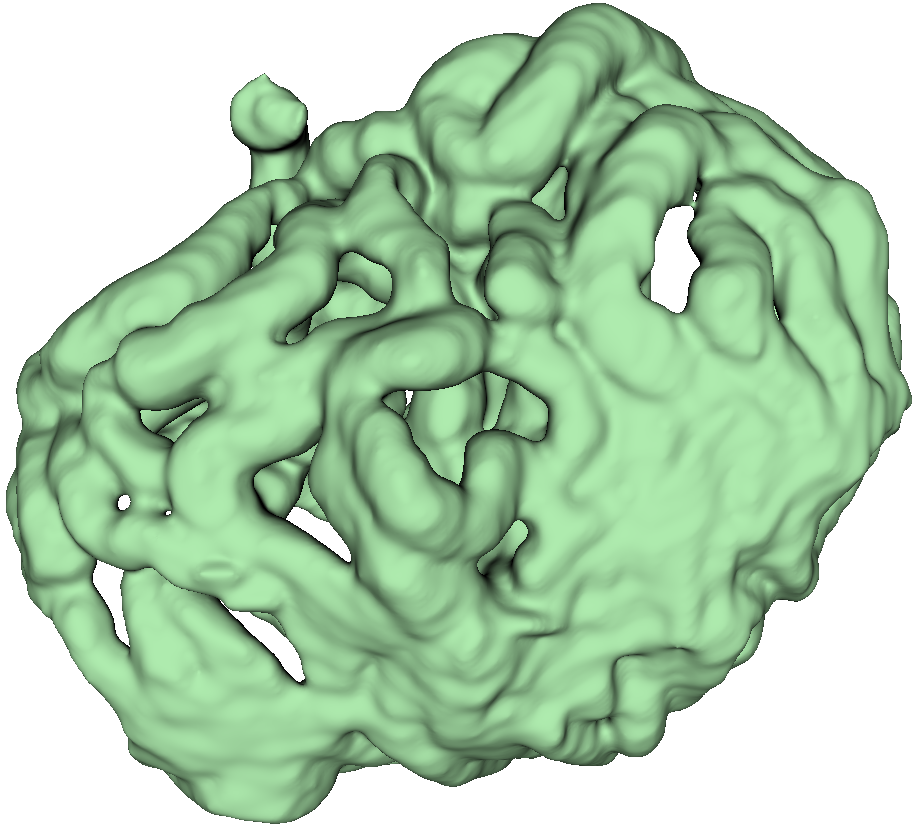}}
	\subfloat[]{\includegraphics[width = 0.4\textwidth]{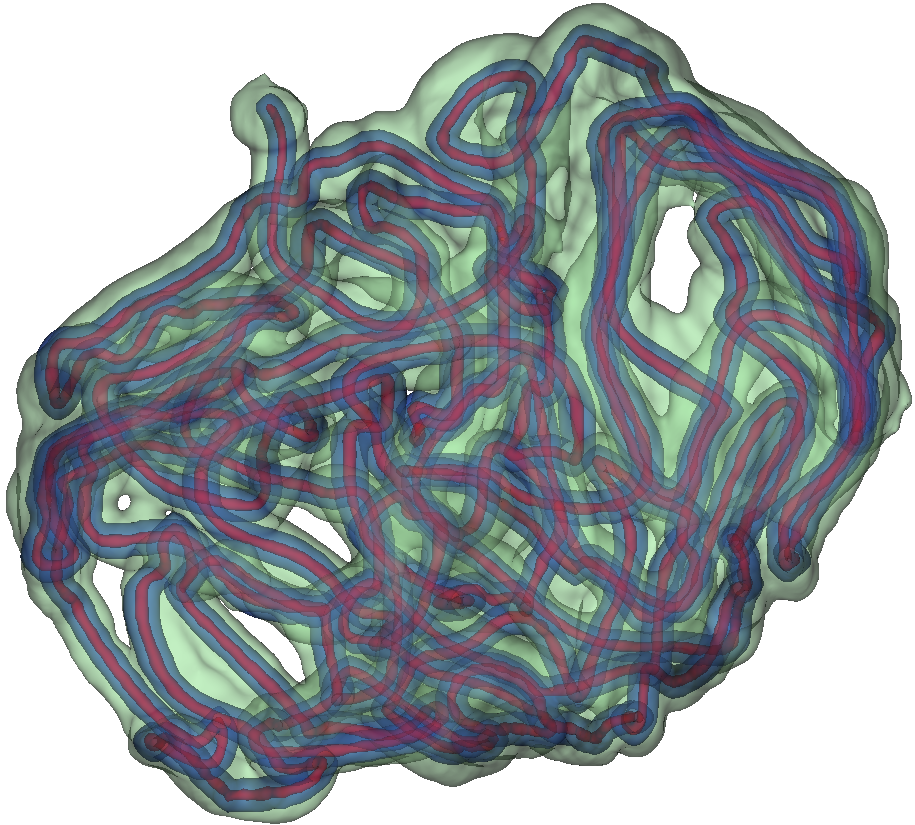}}
	\caption{An example of (a) ground-truth segmentation of the small bowel (green) and (b) corresponding path annotation (red). An inner cylinder (blue) generated by dilating the path annotation, which is used for model training, is also shown.}
	\label{fig:ex_segm_cyl_cl}
\end{figure}

% applying persistent homology for small bowel segmentation 
The small bowel has a cylindrical shape but also has many touchings with different parts along its path, which makes it have variable topologies across patients and time. It is inappropriate to apply a constant topological constraint to such organs. Fig.~\ref{fig:ex_segm_cyl_cl} shows an example of the small bowel path covered by the lumpy ground-truth (GT) segmentation.

% proposed method & contribution
Thus, we present a novel method where a cylindrical topological constraint is applied during network training while addressing the property of variable topologies of the small bowel. We propose to use a network equipped with an additional branch to predict an inner cylinder of the small bowel. The inner cylinder with a smaller diameter still maintains the whole small bowel path but is free of the touching issue. A cylindrical shape constraint applied on this augmented branch guides the network to generate a topologically correct segmentation. The proposed method is evaluated on a dataset composed of high-resolution abdominal CT scans and corresponding dense segmentation GTs, which enables stricter evaluation compared to the previous work using sparse annotation. Four evaluation metrics are used to analyze the results in diverse aspects. Paired t-tests are also conducted to show the statistical significance of the proposed method.

\section{Method}

\subsection{Dataset}

We collected 10 high-resolution abdominal CT scans which were acquired with oral administration of Gastrografin. Compared to the arterial phase CT scans used in \cite{zhang13}, ours are routine contrast-enhanced CT scans which were done during the portal venous phase. All volumes are resampled to have isotropic voxels of $1mm^3$. The original CT scans include a wide range of body regions from chest to pelvis. We manually cropped the abdominal region to include all the small bowel to reduce the computational burden during training.

GT labels were achieved by an experienced radiologist using 3DSlicer\footnote{\url{https://www.slicer.org}}~\cite{fedorov12} based on the following steps. Firstly, the path of the small bowel is drawn as interpolated curves which connect a series of manually placed points inside the small bowel. Secondly, we grow the curves using a margin of $30mm$ and threshold it again using a Hounsfield unit (HU) range of $-80 \sim 200$. Manually drawn air pockets, which have much lower HU values, are also added during the second step. Produced errors are manually fixed in the final step. We note that this annotation procedure took one or two full days for each volume.
%Tracing the small bowel path in a 3D volume is very difficult due to indistinct bowel walls.
Finally, the dataset includes the two types of labels, which are the path and segmentation of the small bowel as shown in Fig.~\ref{fig:ex_segm_cyl_cl}.

\subsection{Persistent Homology}

PH is an emerging tool in topological data analysis, where how long each topological feature of different dimensions persists is measured as some threshold changes (filtration). Data types that can be studied with PH include point clouds, networks, and images~\cite{otter17}. Various types of complexes are constructed from given data, and the computation of PH is conducted on top of it. Cubical complex is a more natural representation for image data than a simplicial complex. In this section, we focus on explaining how PH is involved in our method. We refer the reader to \cite{otter17} for detailed information on PH. 

Voxels in a 3D volume compose elementary cubes in a cubical complex. Considering a predicted probability image $Y$ of an input image $X$, we could consider a super-level set $S_{Y}(p)$ of $Y$, which is the set of voxels for which the probability value is above some threshold value $p$. Decreasing $p$ from 1 to 0 makes a sequence of growing $S_{Y}(p)$'s:

\begin{equation}
	\label{eq:filtration}
	%\emptyset \subseteq S_{Y}(1) \subseteq S_{Y}(p_{1}) \subseteq S_{Y}(p_{2}) \subseteq ... \subseteq
	S_{Y}(1) \subseteq S_{Y}(p_{1}) \subseteq S_{Y}(p_{2}) \subseteq ... \subseteq S_{Y}(0).
\end{equation}

When $S_{Y}(p)$ grows, topological features, e.g., connected components, loops, hollow voids, etc., are created and destroyed as new voxels join. The life time of each feature is recorded as a bar, and a set of bars, called the barcode diagram, is achieved. The barcode diagram is exemplified in Fig.~\ref{fig:res_ph_barcode}. We denote the birth and death values of the $i$-th longest bar of dimension $k$ as $b_{k,i}$ and $d_{k,i}$, respectively. The Betti number $\beta_{k}$ counts the number of topological features of dimension $k$ in $S_{Y}(p)$. $\beta_{0}$, $\beta_{1}$, $\beta_{2}$ are the numbers of connected components, loops or holes, and hollow voids, respectively. Since the segmented inner cylinder of the small bowel should be all connected and has no loops or cavities within it, the values should be 1, 0, and 0.

As mentioned, the computation of PH is differentiable, thus can be incorporated into the training process of neural networks~\cite{clough19}. We introduce the topological loss used in \cite{clough19} since it applies in our method as follows:

\begin{equation}
	\label{eq:topo_loss_kth}
	L_{k}(\beta_{k}^{*}) = \sum_{i=1}^{\beta_{k}^{*}}(1-|b_{k,i}-d_{k,i}|^2) + \sum_{i=\beta_{k}^{*}+1}^{\infty}|b_{k,i}-d_{k,i}|^2,
\end{equation}

\begin{equation}
	\label{eq:topo_loss_sum}
	L_{topo} = \sum_{k}L_{k}(\beta_{k}^{*}),
\end{equation}

Where $\beta_{k}^{*}$ denotes the desired Betti numbers of the target object. Since $\beta_{k}^{*}$ is given as prior knowledge, the loss requires no supervision. It is minimized when the computed barcode diagram has only $\beta_{k}^{*}$ bars of length 1 for each dimension $k$. Therefore, the desired topology, which is specified by $\beta_{k}^{*}$, is reflected into the resulting $Y$ after training. 

In the post-processing framework used in \cite{clough19}, a network $f$ trained using a supervised training set and a supervised loss is again fine-tuned for each test image with the topological loss. Since the individual fine-tuning for each test image is demanding in test time, we fine-tune the network using the training set with the loss function as follows:

\begin{equation}
	\label{eq:total_loss}
	L(X;w,w') = \frac{1}{V}|f(X,w)-f(X,w')|^{2} + \lambda L_{topo}(X,w'),
\end{equation}

Where $V$ is the number of voxels in the volume, and $\lambda$ is the weight for $L_{topo}$. This achieves a new set of weights $w'$, which reflects the desired topology, while minimizing the change from the pretrained weights $w$.

\subsection{Network}

\begin{figure}[t]
	\centering
    \includegraphics[width = 1\textwidth]{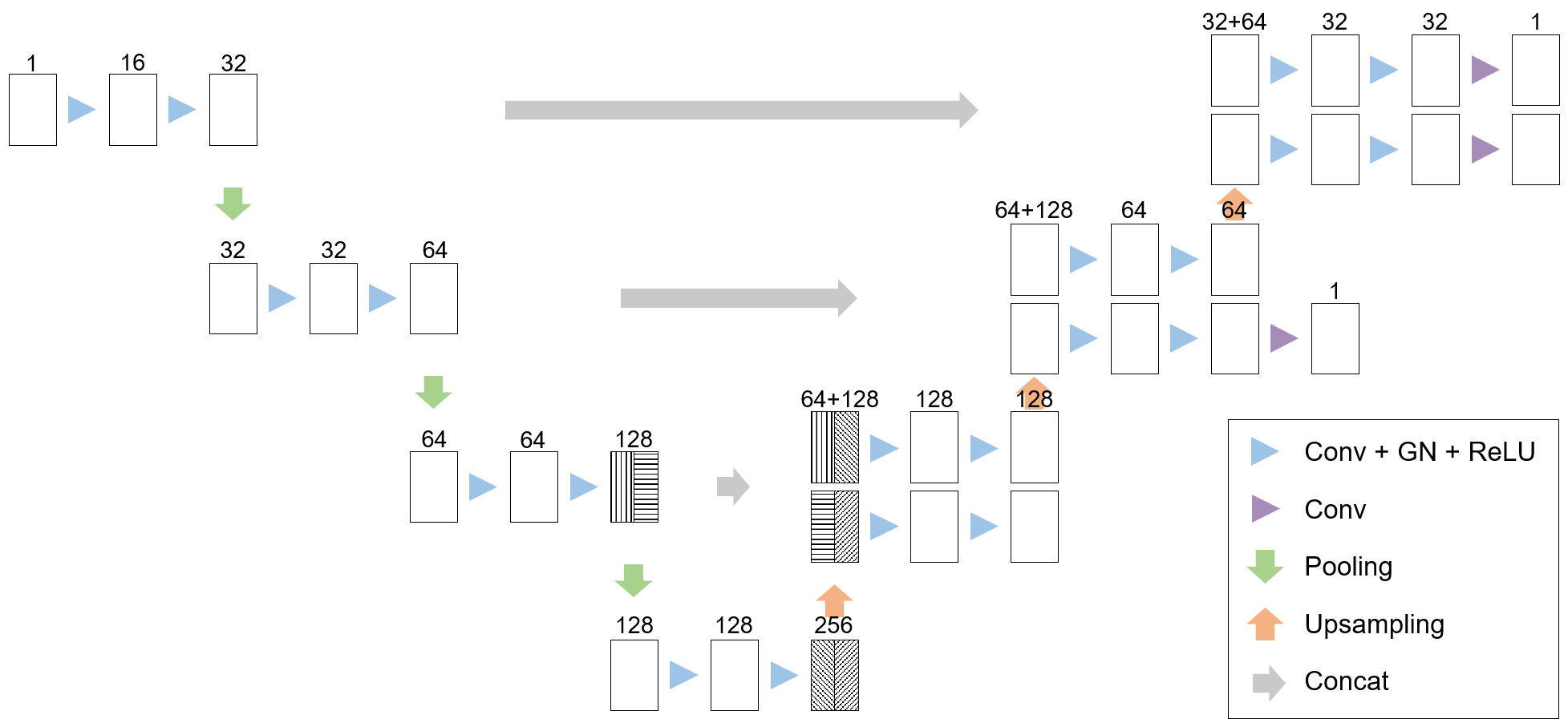}
	\caption{Network architecture for the proposed method. Boxes represent feature maps. The number of channels is denoted on top of each feature map. GN represents the group normalization~\cite{wu18}. The network has two separate decoder paths for segmentation and inner cylinder prediction. While the lower level features of the encoder are mutually used in the separate decoders by skip connections, the high level features are evenly divided and separately used for each decoder in order to generate separate task-specific features without modifying the encoder structure. The decoder path for inner cylinder prediction has one side output layer which enables the application of persistent homology at the smaller spatial resolution.}
	\label{fig:network}
\end{figure}

Fig.~\ref{fig:network} shows the network architecture for the proposed method, which is based on the structure of the 3D U-Net~\cite{cicek16}. The differences are as follows: 1) Our network has two separate decoder paths for segmentation and inner cylinder prediction. The inner cylinder with a smaller diameter still maintains the whole small bowel path but is free of the touching issue. During network training, we apply a cylindrical topological constraint on the inner cylinder prediction instead of the segmentation. 2) The decoder path for inner cylinder prediction has one side output layer which enables the application of PH at the smaller spatial resolution. We found that the PH computation on the full resolution of a 3D volume is prohibitive in both terms of memory and time. Applying the topological loss (Eq.~\ref{eq:topo_loss_sum}) onto the smaller spatial resolution is therefore necessary.

%The added parts of the network give us a chance of regularizing the network training using the cylindrical topological constraint of the small bowel under consideration on its practicality. The parts are detachable in test time, thus do not increase the required resource.
The added parts of the network enable the correct application of the cylindrical topological constraint for small bowel segmentation during training. Those are detachable in test time, thus do not increase the required resource.

\subsection{Evaluation Details}

We used a NVIDIA Tesla V100 32GB GPU to conduct experiments. The mini-batch size was set as $1$, and the CT volumes were resized with a scale factor of $1/2$ to make the network training fit in the GPU memory. The inner cylinder diameter was set as $3$ voxels by experiments. We used the generalized Dice loss~\cite{sudre17} for the initial supervised training.

We implemented the network using PyTorch, and used the PyTorch-compatible layers introduced in \cite{bruel19} to compute the PH and its gradients. For training the network, we used an AdamW optimizer~\cite{loshchilov19} and a weight decay of $5 \times 10^{-4}$. The learning rate of $10^{-4}$ was used for all experiments. We used $0.01$ for $\lambda$.

We perform a 2-fold cross validation and use Dice coefficient, Hausdorff distance (HD), 95\% Hausdorff distance (HD95), and average symmetric surface distance (ASD) as our evaluation metrics. Paired t-tests are also conducted to show the statistical significance of the proposed method.

\section{Results}

\subsection{Quantitative Evaluation}

Table~\ref{tab:quan_res} provides quantitative results of different segmentation models. The proposed method, `Seg + Cyl + PH', shows clear improvements for all metrics. The relative improvements are +1.75\% in Dice, -12.31\% in HD, -24.25\% in HD95, and -14.77\% in ASD, compared to the baseline, `Seg'. Since our topological constraint takes effect globally, the improvements are more obvious with the surface distance measures than Dice. The improvements are not from augmenting the network itself with an additional branch since there is no clear improvement for `Seg + Cyl'. Also, a comparable method where the topological constraint is directly applied on the segmentation, `Seg + PH', shows rather worse results. This confirms the need for our sophisticated scheme for topologically better small bowel segmentation. The p-values computed by conducting paired t-tests between the baseline method and the other methods with the Dice coefficients show the statistical significance of the proposed method.

\begin{table}
\caption{Quantitative comparison of different segmentation models. All models are variants of that in Fig.~\ref{fig:network}. `Seg' denotes a single decoder segmentation network trained using a supervised loss; `Seg + PH' denotes fine-tuning a pretrained segmentation network using the topological loss, but directly on the segmentation outputs; `Seg + Cyl' denotes the network in Fig.~\ref{fig:network}, which is trained using only supervised losses; `Seg + Cyl + PH' denotes the proposed method. For every metric, the mean and standard deviation are presented. Refer to the text for the explanation on the evaluation metrics. P-values are computed by conducting paired t-tests between the baseline method and the other methods with the Dice coefficients.}\label{tab:quan_res}
\begin{tabular}{c|c|c|c|c|c}
\\
Method & Dice & HD (mm) & HD95 (mm) & ASD (mm) & p-value \\
\hline
Seg~\cite{cicek16} & 0.838 $\pm$ 0.044 & 58.894 $\pm$ 20.423 & 16.081 $\pm$ 6.886 & 2.733 $\pm$ 0.879 & -\\
\hline
Seg + PH~\cite{clough19} & 0.822 $\pm$ 0.061 & 64.754 $\pm$ 28.825 & 14.666 $\pm$ 6.612 & 2.802 $\pm$ 1.076 & 0.205\\
\hline
Seg + Cyl & 0.839 $\pm$ 0.048 & 63.968 $\pm$ 22.219 & 14.192 $\pm$ 5.494 & 2.644 $\pm$ 0.840 & 0.679\\
\hline
Seg + Cyl + PH & \textbf{0.852} $\pm$ 0.045 & \textbf{51.642} $\pm$ 9.292 & \textbf{12.180} $\pm$ 6.232 & \textbf{2.330} $\pm$ 0.764 & 0.032\\
\end{tabular}
\end{table}

\subsection{Qualitative Evaluation}

\begin{figure}[t]
	\centering
	\begin{minipage}{1\textwidth}
        \subfigimg[width=0.5\textwidth,pos=ll, font=\color{white}]{A}{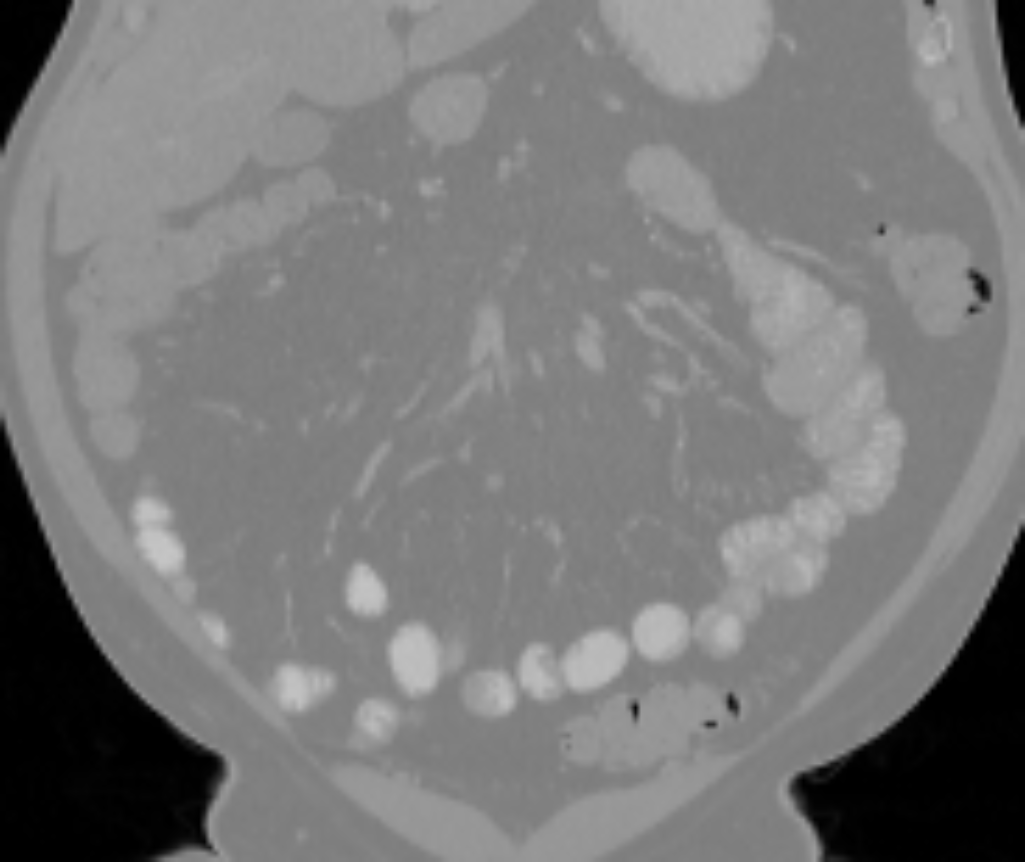}
        \hspace{-0.2cm}
        \subfigimg[width=0.5\textwidth,pos=ll, font=\color{white}]{B}{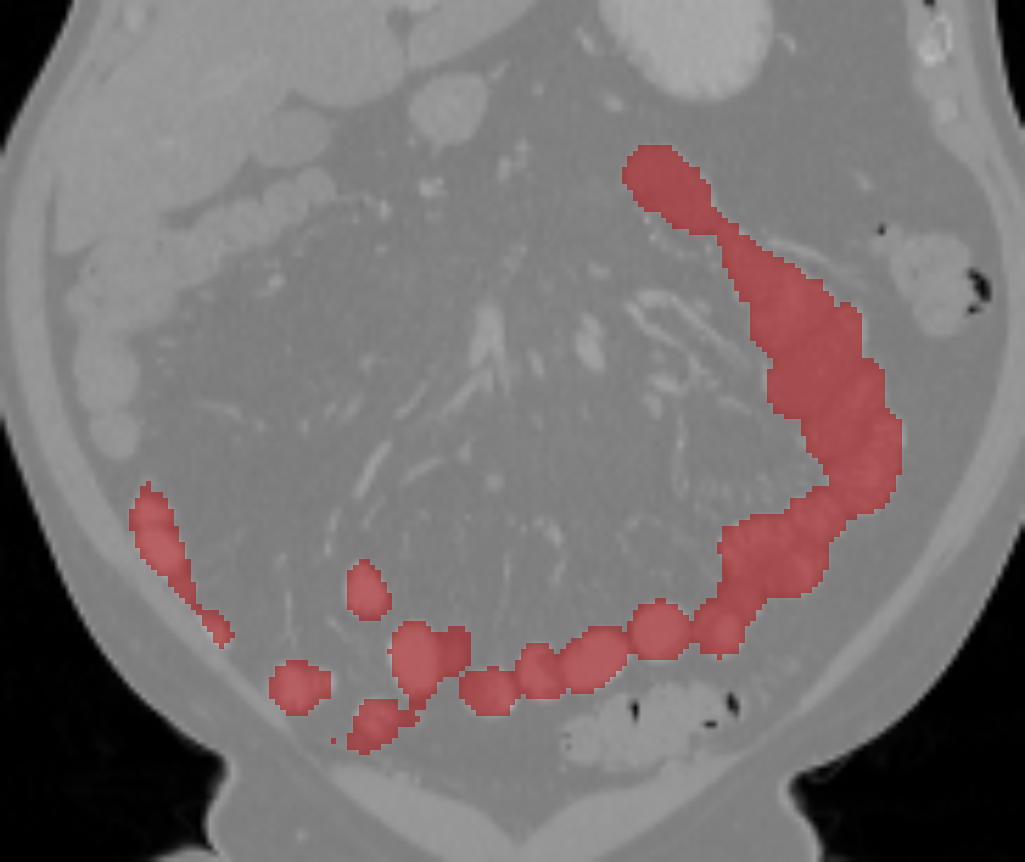}
    \end{minipage}
    \\
	%\vspace{-0.05cm}
    \begin{minipage}{1\textwidth}
        %\subfigimg[width=0.5\textwidth,pos=ll, font=\color{white}]{(c)}{res_2d_seg.png}
        \subfigimg[width=0.5\textwidth,pos=ll, font=\color{white}]{C}{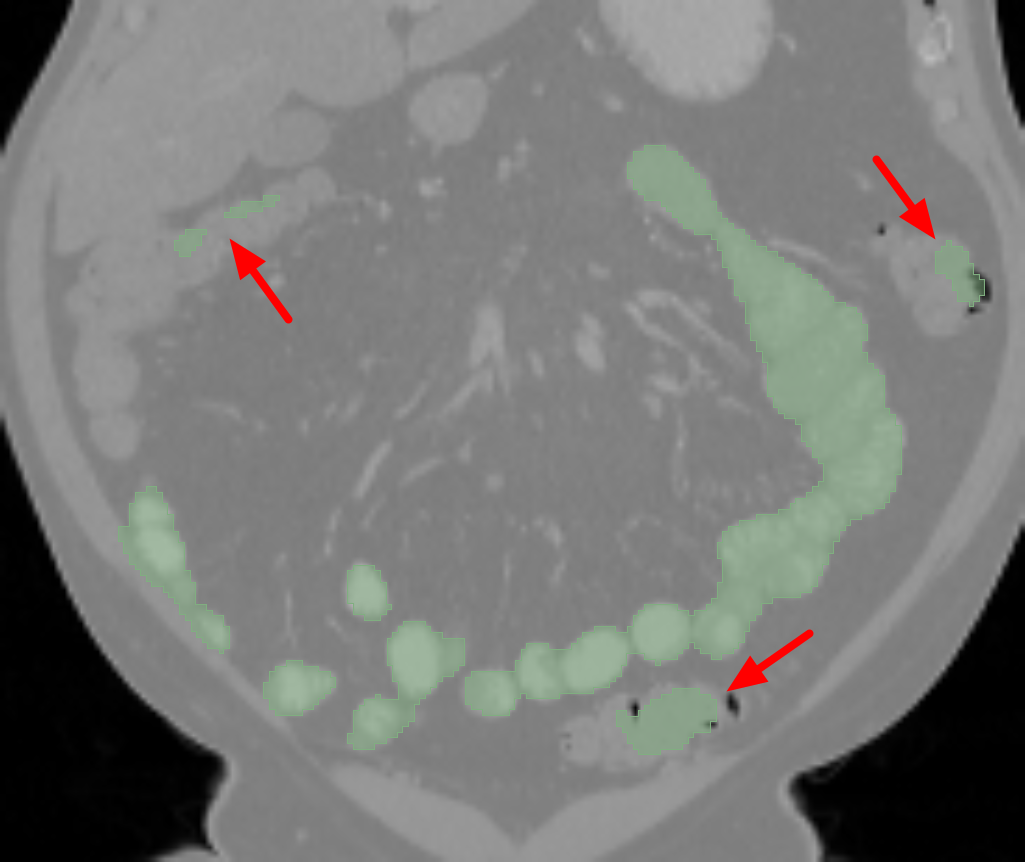}
        \hspace{-0.2cm}
        \subfigimg[width=0.5\textwidth,pos=ll, font=\color{white}]{D}{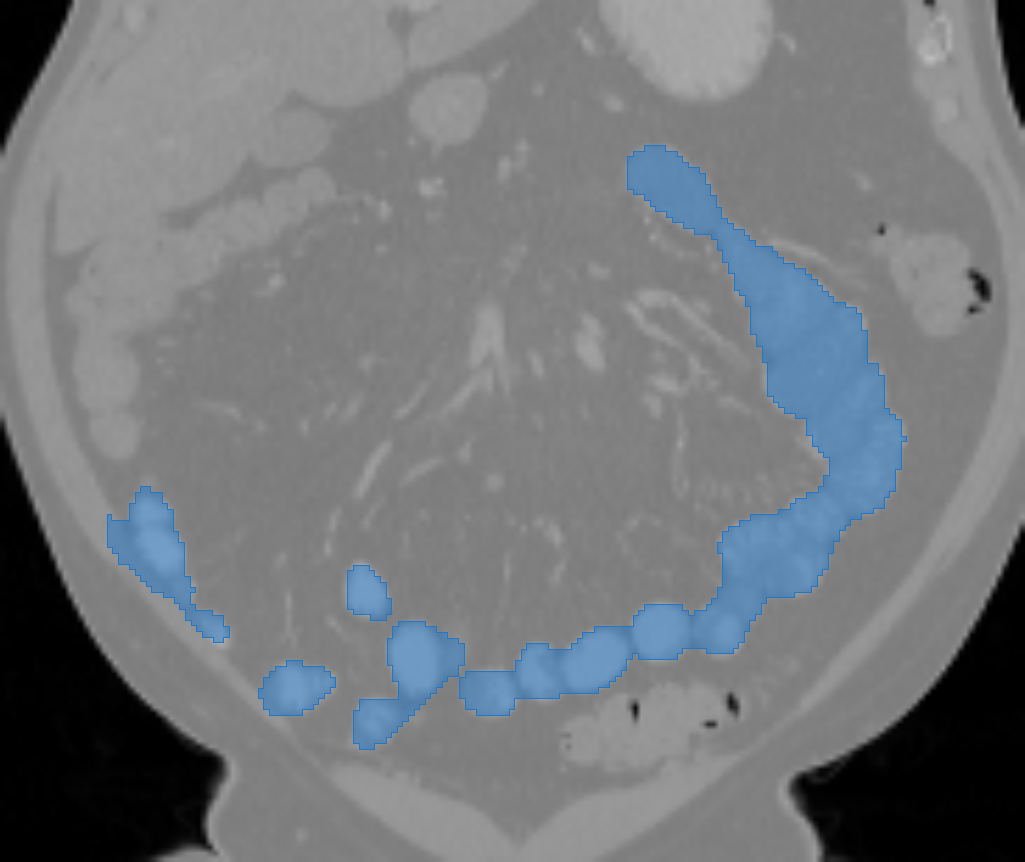}
    \end{minipage}
	\caption{Example segmentation results in coronal view. The images are: (A) an image slice of the input volume, (B) ground-truth segmentation, (C) result of the 3D U-Net~\cite{cicek16}, which corresponds to `Seg' in Table~\ref{tab:quan_res}, and (D) result of the proposed method. Red arrows indicate false positives.}
	\label{fig:res_2d}
\end{figure}

\begin{figure}[!h]
	\centering
	\subfloat[]{\includegraphics[width = 0.33\textwidth]{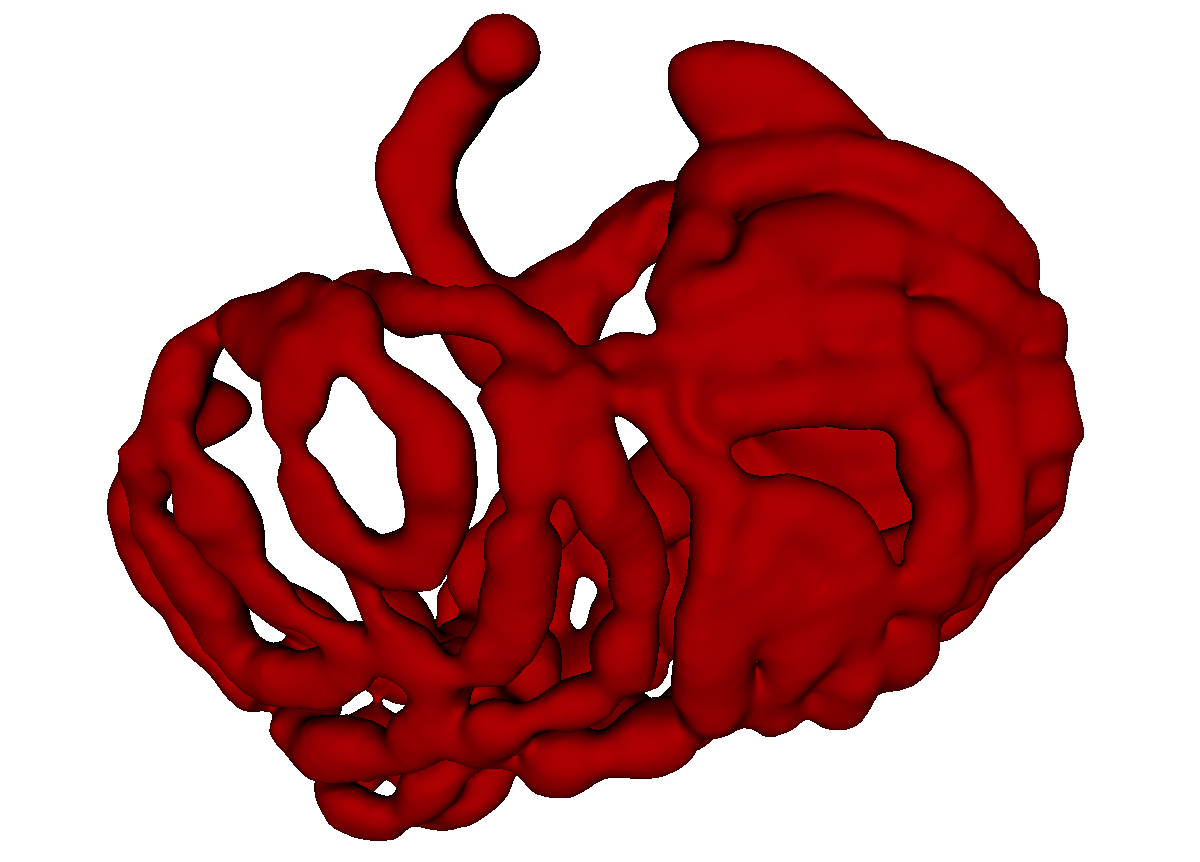}}
	\subfloat[]{\includegraphics[width = 0.33\textwidth]{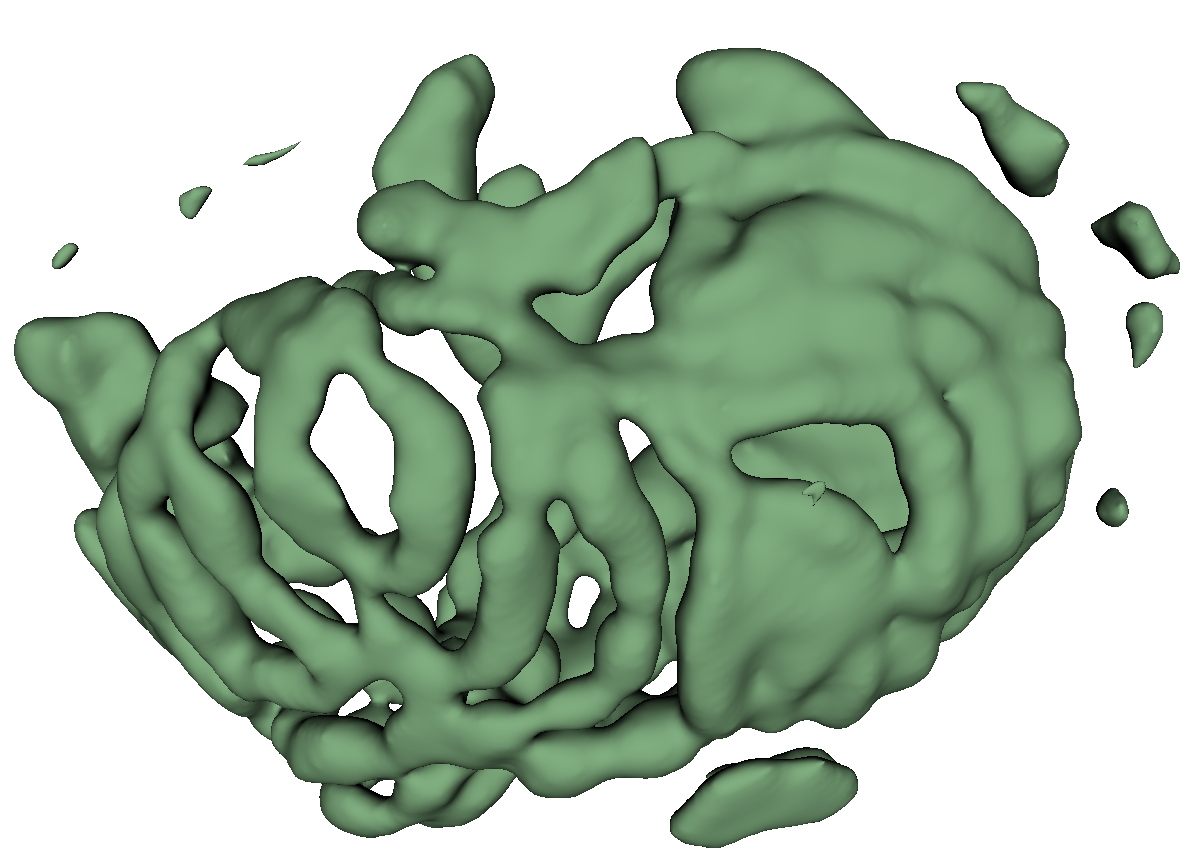}}
    \subfloat[]{\includegraphics[width = 0.33\textwidth]{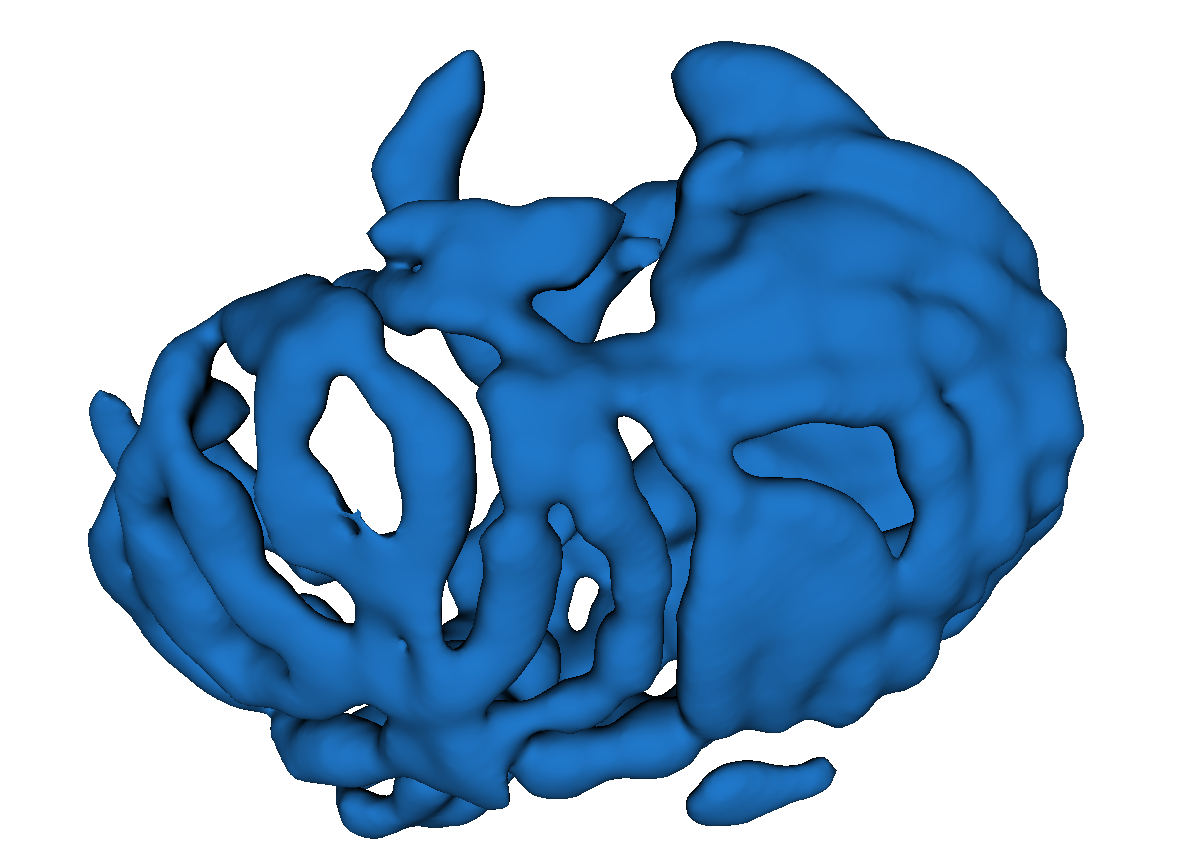}}
	\caption{Example segmentation results in 3D. (a) Ground-truth segmentation. (b) Result of the 3D U-Net~\cite{cicek16}, which corresponds to `Seg' in Table~\ref{tab:quan_res}. (c) Result of the proposed method.}
	%\caption{Example segmentation results in 3D. The images are, from left to right, ground-truth segmentation, result of the 3D U-Net~\cite{cicek16}, which corresponds to `Seg' in Table~\ref{tab:quan_res}, and result of the proposed method.}
	\label{fig:res_3d}
\end{figure}

\begin{figure}[!h]
	\centering
	\subfloat[]{\includegraphics[width = 0.33\textwidth]{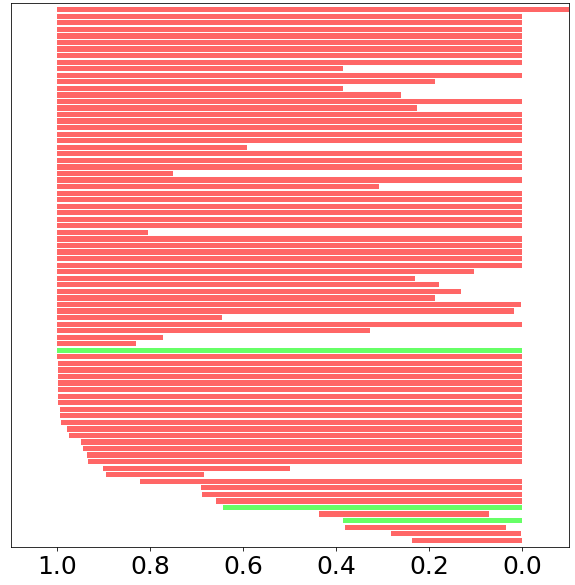}}
	\subfloat[]{\includegraphics[width = 0.33\textwidth]{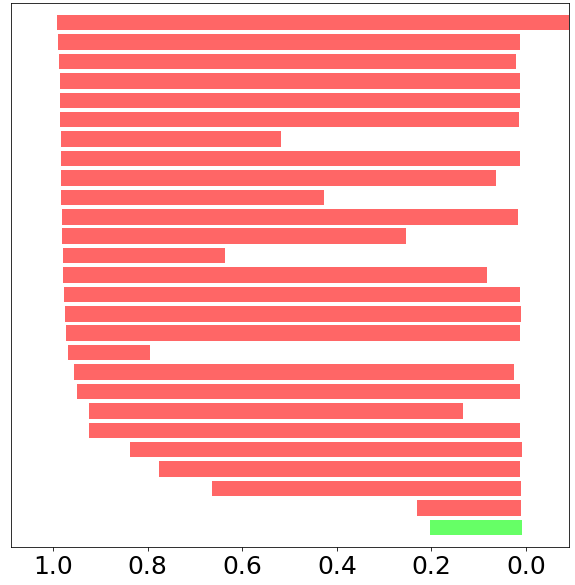}}
	\caption{Example barcode diagrams of (a) the result of the 3D U-Net~\cite{cicek16} and (b) the result of the proposed method. The horizontal axis indicates the filtration values. The left and right ends of each bar represent the birth and death values of the corresponding feature. Red and green colors denote 0- and 1-dimensional features, respectively. 2-dimensional features, which were also considered, are not shown here due to their short life time. }
	\label{fig:res_ph_barcode}
\end{figure}

Fig.~\ref{fig:res_2d} shows example segmentation results. The proposed method eliminates false positives on the large bowel by the help of the applied topological constraint. Fig.~\ref{fig:res_3d} further clarifies the effectiveness of the proposed method by presenting 3D rendered segmentations. The proposed method produces a more topologically correct segmentation of the small bowel with fewer false positives. Fig.~\ref{fig:res_ph_barcode} presents example barcode diagrams, which again show the reduced numbers of connected components and holes within the segmentation result of the proposed method.

\section{Conclusion}

Although PH has been proved effective in several medical tasks and target objects, it requires technical improvement in order to broaden its applicable scope to more complex targets. In this work, we showed how it would be incorporated for improved segmentation of the small bowel which has variable configuration. We acquired dense segmentation GTs, which enabled stricter evaluation of the proposed method. The improved segmentation could help precise localization of diseases, such as inflammatory bowel disease and carcinoid, and preoperative planning by better visualization. In future work, we plan to include CT scans with diverse constrast media to improve the clinical practicality of the proposed method.

\subsubsection*{Acknowledgments.} This research was supported by the National Institutes of Health, Clinical Center and National Cancer Institute.

%
% ---- Bibliography ----
%
% BibTeX users should specify bibliography style 'splncs04'.
% References will then be sorted and formatted in the correct style.

%\bibliographystyle{splncs04}
%\bibliography{miccai20_ref}

\end{document}